\documentclass[aps,pra,twocolumn]{revtex4}
\usepackage{graphicx}

\begin{document}

\title{Probabilistic Quantum Gates between Remote Atoms \\
through Interference of Optical Frequency Qubits}

\author{L.-M. Duan, M. J. Madsen, D. L. Moehring, P. Maunz, R. N. Kohn Jr., and C. Monroe}
\address{FOCUS Center and Michigan Center for Theoretical Physics \\
Department of Physics, University of Michigan, Ann Arbor, MI 48109}

\begin{abstract}
We propose a scheme to perform probabilistic quantum gates on remote trapped
atom qubits through interference of optical frequency qubits. The method does not
require localization of the atoms to the Lamb-Dicke limit, and is not
sensitive to interferometer phase instabilities. Such probabilistic
gates can be used for scalable quantum computation.
\end{abstract}

\maketitle

Entangled quantum states, at the heart of quantum information processing,
are notoriously difficult to generate and control. Generating entangled states becomes dramatically simpler when the entanglement operations are allowed to succeed with only a finite (perhaps small) probability, as long as it is known when the operations
succeed \cite{1,2,3,4,4b}. If entangling \textit{gates} can be implemented in such a
probabilistic fashion, it has recently been shown that scalable quantum
computation is still possible, no matter how small
the gate success probability \cite{5,6}. Compared with
deterministic gates, the additional overhead in resources (such as
the number of qubit manipulations) for probabilistic
quantum computation scales only polynomially with both the size of the
computation and the inverse of the gate success probability
\cite{5}.

There have been recent proposals for implementation of
probabilistic gates \cite{6,7,8}, using atomic qubits inside
optical cavities. In this paper, we propose a new scheme for
probabilistic quantum gate operations that act on trapped atoms or
ions in free space (with or without cavities). Compared with
previous methods, this scheme has two outstanding features.
First, optical frequency qubits are used to connect and entangle
matter qubits at distant locations. The two states comprising this
optical qubit have the same polarization, but differ in frequency
by atomic qubit splitting (typically in the microwave region for
hyperfine atomic qubits). These closely-spaced frequency
components have basically zero dispersion in typical optical
paths, thus this optical qubit is highly insensitive to phase
jitter inherent in optical interferometers. Such optical frequency
qubits have been demonstrated in a very recent experiment
\cite{exp}, through control of a trapped cadmium ion with
ultrafast laser pulses. Second, the proposed entangling scheme
does not require localization of the atoms to the Lamb-Dicke
limit.  Motion of the atomic qubits can be larger than the optical
wavelength. Although ions can be localized well under the
Lamb-Dicke limit through laser cooling in a strong trap, the
elimination of this stringent requirement should greatly simplify
experiments. This is particularly important for ions confined in
miniature electrode structures fabricated on a chip
\cite{exp2,surfacetrap}, where ion heating may become more
significant \cite{deslauriers}. This feature is also crucial for
neutral atom qubits, where confinement to the Lamb-Dicke regime is
very difficult.

Our scheme is illustrated in Fig. 1. The qubit is represented by two
$S_{1/2}$ ground state hyperfine levels of an alkali-like atom (ion), with $\left|
0\right\rangle \equiv \left| F,m=0\right\rangle $, \ and $\left|
1\right\rangle \equiv \left| F+1,m=0\right\rangle $. These ``clock" states are
particularly insensitive to stray magnetic fields. In the figure, for
simplicity, we take $F=0$, which is the case for ions such as $^{111}$Cd$^+$,
but the scheme works for any value of $F$. To perform a
probabilistic gate on two remote atoms 1 and 2, we first excite both of the
atoms to the $P_{1/2}$ excited electronic state with a $\pi $-polarized ultrafast laser pulse
\cite{note1}. We assume the laser has a bandwidth which is larger than the
hyperfine splitting ($14$ GHz for $^{111}$Cd$^{+}$), but smaller than the
fine structure splitting between $P_{1/2}$ and $P_{3/2}$ ($74$ THz for $%
^{111}$Cd$^{+}$). Typical picosecond pulses used in experiments
(bandwidth $\sim 500$\ GHz) satisfy these requirements \cite{exp}.
Under the above condition, we can assume the pulse only drives the
D1 transition from the ground state $S_{1/2}$ to the excited state
$P_{1/2}$ \cite{note2}. Due to dipole selection rules, for a $\pi
$-polarized pulse, only the hyperfine transitions $\left|
F,m=0\right\rangle \rightarrow \left| F^{\prime
}+1,m=0\right\rangle $ and $\left| F+1,m=0\right\rangle
\rightarrow \left| F^{\prime },m=0\right\rangle $ are allowed,
where the upper hyperfine spin $F^{\prime }=F$. Thanks to the
selection rules, each qubit state is transferred to a unique
excited hyperfine level after the pulsed laser excitation. This
point is critical for successful gate operation.

\begin{figure}[tbp]
\includegraphics[height=5cm,width=8cm]{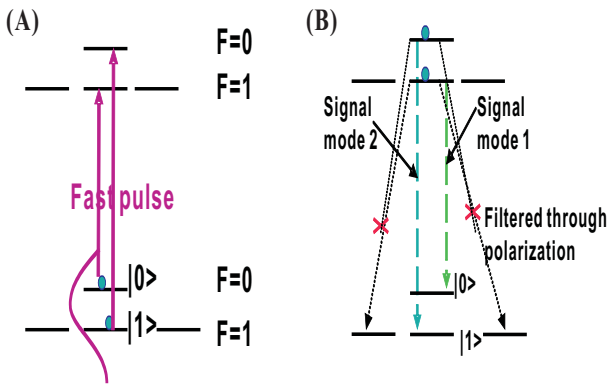}
\caption[Fig.1 ]{The atomic level configuration and the laser
excitation scheme. (A) An ultrafast laser pulse transfer the
atomic qubit state from the ground levels to the excited levels.
(B) The atom decays back to the ground levels, with the frequency
of the spontaneously emitted photon correlated with the atomic
qubit state (marked as the signal mode 1 and 2 in the figure). The
photon from the $\sigma ^{\pm }$ decay channels is filtered
through polarization selection. }
\end{figure}

After this laser excitation, the atoms eventually decay back to
their ground $S_{1/2}$ states. There are several decay channels, through the
emission of either $\pi$-polarized or $\sigma ^{\pm }$-polarized
spontaneous emission photons (see Fig. 1b). We first consider the decay channels with a $\pi $%
-polarized emission photon. In this case, the excited levels $\left|
F^{\prime }+1,m=0\right\rangle $ and $\left| F^{\prime },m=0\right\rangle $
can only decay back to the ground states $\left| F,m=0\right\rangle $ and $%
\left| F+1,m=0\right\rangle $, respectively. While photons from these two decay
channels have the same polarization, they have slightly different
frequencies. The frequency difference is given by $\Delta _{HF}^{S}+\Delta
_{HF}^{P}$, the sum of the hyperfine splittings of the ground $S_{1/2}$ and excited
$P_{1/2}$ states. This frequency difference is typically
much larger than the natural linewidth of the excited level \cite{note3}, so
the corresponding photons from the two $\pi $-decay channels are well-resolved
in frequency. This defines two frequency modes for the emitted
photon field, and we call them $\nu _{0}$ and $\nu _{1}$ modes,
respectively. If the atom is initially in the qubit state $\left| \Psi
_{a}\right\rangle =c_{0}\left| 0\right\rangle +c_{1}\left| 1\right\rangle $,
then after this excitation-decay process the atom-photon system evolves to an
entangled state
\begin{equation}
\left| \Psi _{ap}\right\rangle =c_{0}\left| 0\right\rangle \left| \nu
_{0}\right\rangle +c_{1}\left| 1\right\rangle \left| \nu _{1}\right\rangle
\end{equation}
if we only collect the photon from the $\pi $ decay channels,
where $\left| \nu _{0}\right\rangle $ and $\left| \nu
_{1}\right\rangle $ represent a single photon state in the
frequency modes $\nu _{0}$ and $\nu _{1}$, respectively. Any
photon from the $\sigma ^{\pm }$ decay channels is assumed to be
blocked through a polarization filter. This result is somewhat
similar to the previous demonstration of the atom-photon
entanglement \cite {4,4b}, but there are important differences.
First, the final state $\left| \Psi _{ap}\right\rangle$ keeps
track of the information $c_{0},c_{1}$ of the initial qubit state.
Thus, the scheme here is not just an entangling protocol
\cite{10}, but is instead an entangling \textit{gate} with the
final quantum state depending on the initial state. As we will see
later, this type of gate can form the basis for scalable quantum
computation, and is therefore more powerful than merely an
entangling operation. Second, the spontaneous emission photon with
either frequency $\nu _{0}$ or $\nu _{1}$ has the same spatial
mode, so good spatial mode-matching of this photonic qubit is
possible even if we increase the solid angle of collection. In the
previous entangling protocol \cite{10,4,4b}, the quantum
information is carried by different polarization modes of the
photon, which have different spatial emission patterns.  This
requires small collection solid angles in order to both maintain
orthogonality and ensure adequate spatial matching of the photonic
qubit states.

To perform a gate on two remote atoms, the
spontaneous emission photons from the decay channels in each atom are collected in a
certain solid angle, and directed onto a beam splitter for
interference (see Fig. 2). The output of the beam splitter is
measured by two single-photon detectors. We keep the resulting outcome
atomic state only when we register a photon from each detector. In
this case, what we have performed is a ``measurement gate"
on the atoms 1 and 2. It corresponds to a quantum non-demolition
measurement of the
operator $Z_{1}Z_{2}$, where $Z_{i}$ (or $X_{i}$) stands for the $z$ (or $x$%
) component of the Pauli matrix associated with atomic qubit $i$. After the
coincidence measurement of photons on both detectors, the atomic state is projected to the
eigenspace of $Z_{1}Z_{2}$ with $-1$ eigenvalue. To see this, we note that before the measurement, the state of both atom-photon systems can be written as $\left| \Psi _{ap}\right\rangle _{1}\otimes
\left| \Psi _{ap}\right\rangle _{2}$, where $\left| \Psi _{ap}\right\rangle
_{1}$ has the form of Eq. (1), and $\left| \Psi _{ap}\right\rangle _{2}$ can
be written as $\left| \Psi _{ap}\right\rangle _{2}=d_{0}\left|
0\right\rangle _{2}\left| \nu _{0}\right\rangle _{2}+d_{1}\left|
1\right\rangle _{2}\left| \nu _{1}\right\rangle _{2}$. To register a photon
from each detector, the two photons before the beam splitter need to go to
different sides, which means they should be in the anti-symmetric component $%
\left| \Phi _{AS}\right\rangle =\left( \left| \nu
_{0}\right\rangle _{1}\left| \nu _{1}\right\rangle _{2}-\left| \nu
_{1}\right\rangle _{1}\left| \nu _{0}\right\rangle _{2}\right)
/\sqrt{2}$ (for photons in the symmetric states, they always go to
the same detector). So, given that the photons take separate paths
after the beam-splitter, the state of the atoms 1,2 is given by
the projection
\begin{eqnarray}
\left| \Psi _{12}\right\rangle  &\propto &\left\langle \Phi _{AS}\right|
\left| \Psi _{ap}\right\rangle _{1}\otimes \left| \Psi _{ap}\right\rangle
_{2}  \nonumber \\
&\propto &c_{0}d_{1}\left| 0\right\rangle _{1}\left|
1\right\rangle
_{2}-c_{1}d_{0}\left| 1\right\rangle _{1}\left| 0\right\rangle _{2} \\
&\propto &Z_{1} \left( I-Z_{1}Z_{2}\right) \left| \Psi
_{a}\right\rangle _{1}\otimes \left| \Psi _{a}\right\rangle _{2},
\nonumber
\end{eqnarray}
where $I-Z_{1}Z_{2}$ is the corresponding projector, and $Z_{1}$
is a trivial additional single-bit gate on atom 1 which we will
neglect in the following. This measurement gate, of course, only
succeeds with a finite probability. The overall success
probability is given by $p_{s}=\eta _{d}^{2}\eta _{c}^{2}\eta
_{b}^{2}/4$, where $\eta _{d}$ is the quantum efficiency of each
detector, $\eta _{c}$ is the photon collection efficiency
(proportional to the solid angle), and $\eta _{b}$ is the
branching ratio for the atom to decay along the $\pi $ channel. We
have an additional factor of $1/4$ in $p_{s}$ describing the
average probability for the two spontaneous emission photons to go
to different detectors (averaged over all the possible initial
atomic states). In the above contributions to the success
probability, the collection efficiency is typically the smallest
and thus dominates the overall efficiency. That is why is
important to increase the collection solid angle as much as
possible. Alternatively, one can also increase this efficiency
with the use of optical cavities surrounding the atoms
\cite{CQED}.

\begin{figure}[tbp]
\includegraphics [height=4cm,width=8cm]{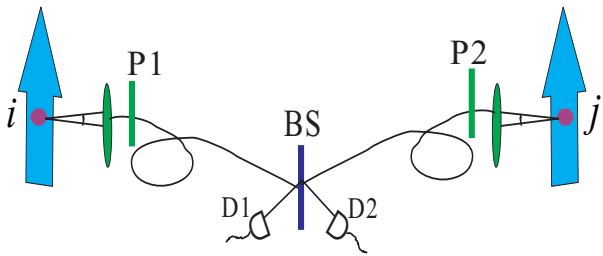}
\caption[Fig.2 ]{The ZZ measurement gate on the atoms i and j. The
spontaneous emission photons from the $\pi$ decay channels of
these two atoms are collected, interfered at the beam splitter
(BS), and then detected by two single-photon detectors (D1 and
D2). If each detector registers a photon, the atomic state is
projected onto the eigen-space of the $Z_iZ_j$ operator.}
\end{figure}

The above measurement gate is robust to noise. We do not require that the atoms
be localized to the Lamb-Dicke limit. In general, atomic motion occurs
with a time scale of the trap frequency $\nu _{t}$, typically much
smaller than the decay rate $\gamma$ of the excited atomic level. Thus,
for each spontaneous emission pulse, we can safely assume the atom to be in
a fixed but random position $\mathbf{r}$. In this case, both of the
frequency components $\left| \nu _{0}\right\rangle $
and $\left| \nu _{1}\right\rangle $ will acquire the same random phase factor
proportional to $e^{i\mathbf{k\cdot r}}$, where $\mathbf{k}$ is the wave
vector associated with the spontaneous emission photon. This overall
phase therefore has no effect on
the resultant measurement gate as shown in Eq. (2). If we take into account
of the motion of the atom within the pulse duration, the pulse from this
moving atom also has a slight Doppler shift $\delta \omega =\mathbf{k\cdot
v\sim }\left| \mathbf{k}\right| \nu _{t}l_{s}$ in its frequency, where $%
\mathbf{v}$ is the random atom velocity at that moment, and $%
l_{s}$\ is the characteristic length scale for the atom oscillation. We need
this random Doppler shift to be significantly smaller than the bandwidth of the
pulse in order to have a good shape matching of the spontaneous emission
pulses from different atoms. So, there is a further requirement $\left|
\mathbf{k}\right| \nu _{t}l_{s}\ll \gamma $, which is consistent with the
assumption $\nu _{t}\ll \gamma $. Finally, this gate is also very
insensitive to the birefringence and the phase drift in the optical
interferometer. Both of the components $\left| \nu _{0}\right\rangle $ and $%
\left| \nu _{1}\right\rangle $ have the same polarization, and they are very
close in frequency. So, they essentially experience the same noisy phase shift
under fluctuation of the optical path length, again cancelling.

We have shown how to perform a probabilistic measurement gate on
remote atoms by projecting the system state to the eigenspace of
the $Z_{1}Z_{2}$ operator. Such a gate only succeeds with a small
probability, but it is very robust to noise. This type of
probabilistic gate can lead to efficient quantum computation, no
matter how small the success probability is \cite {5,6,note4}. The
proof is based on efficient construction of the two-dimensional
(2D) cluster state, which has been shown to be a sufficient
resource for universal quantum computation \cite{11}. To construct
a 2D cluster state of size $n$, the number of pulses (elementary
operations) scales with $n$ by $n\ln n$, and scales with the
inverse of the gate success probability $1/p_{s}$ nearly
polynomially. The exact scaling formula can be found in Ref.
\cite{5}, where it is derived for the case of probabilistic
controlled phase flip (CPF) gates. For the ZZ measurement gate,
the scaling formula is almost the same. To see this, we can simply
note the following two facts: (1) If one starts with two qubits
(atoms) in the co-eigenstate of $X_{1}$ and $X_{2}$ (a product
state), the final state after a ZZ measurement is projected to a
co-eigenstate of the stabilizer operators $Z_{1}Z_{2}$ and
$X_{1}X_{2}$, which is equivalent to the two-qubit cluster state
under single-bit rotations \cite{6}; (2) Assume that one has
prepared two 1D cluster chains, each of $n$ qubits. The stabilizer
operators for the boundary qubits n and n+1 of the two chains are
denoted by $X_{n}Z_{n-1}$ and $X_{n+1}Z_{n+2}$, respectively. A ZZ
measurement of these two boundary qubits generates the new
stabilizer operators $Z_{n}Z_{n+1}$ and
$X_{n}X_{n+1}Z_{n-1}Z_{n+2}$. This operation actually connects the
two chains into a cluster state of $2n-1$ qubits (the central
qubits n and n+1 together represent one logic qubit with the
encoded $X_{L}=X_{n}X_{n+1}$ and $Z_{L}=Z_{n}$ or $Z_{n+1}$). From
these two facts, we can derive the recursion relations. Following
the same argument as in the case for the CPF\ gate \cite{5}, we
can show that started with two cluster chains each of n qubits,
the average length of the
cluster state after this probabilistic measurement gate is given by $%
n^{\prime }=\sum_{i}\left( 2n-1-4i\right) p_{s}\left(
1-p_{s}\right) ^{i}\approx 2n-n_{c}$, where the critical length
$n_{c}=1+4\left( 1-p_{s}\right) /p_{s}$ and $p_{s}$ is the gate
success probability. Compared with the case of the CPF gate, the
only difference is the critical length $n_{c}$ changes from
$4\left( 1-p_{s}\right) /p_{s}$ to $1+4\left( 1-p_{s}\right)
/p_{s}$, and such a change is negligible in the case of a small
success probability with $4/p_{s}\gg 1$. So, for this ZZ
measurement gate, we find nearly the same scaling formula derived in Ref. \cite{5}.

In summary, we have proposed a scheme for probabilistic gates on
remote trapped atoms or ions in free space, based on interference
of optical frequency qubits from the atomic spontaneous emission
driven by ultrafast laser pulses. This gate scheme does not
require localization of the atoms to the Lamb-Dicke limit, and is
robust to practical phase noise in the optical interferometers.
This type of probabilistic gate could lead to alternative way for
efficient quantum computation.

We thank R. Raussendorf, B. Blinov, S. D. Barrett, and P. Kok for
helpful discussions. This work was supported by National Science
Foundation award 0431476 and the ITR program, the National
Security Agency and the Disruptive Technology Organization under
Army Research Office contract W911NF-04-1-0234, and the A. P.
Sloan Foundation..


\begin{thebibliography}{99}
\bibitem{1}  L. M. Duan, M. D. Lukin, J. I. Cirac, P,. Zoller, Nature
\textbf{414}, 413 (2001).


\bibitem{2}  E. Knill, R. Laflamme, and G. Milburn, Nature \textbf{409}, 46
(2001).


\bibitem{3}  A. Kuzmich, \textit{et al.}, Nature \textbf{423}, 731 (2003);
C. H. van der Wal \textit{et al.}, Science \textbf{301}, 196 (2003).


\bibitem{4}  B. B. Blinov, D. L. Moehring, L.-M. Duan, C. Monroe, Nature
\textbf{428}, 153-157 (2004).

\bibitem{4b}  D. L. Moehring, M. J. Madsen, B. B. Blinov, and C. Monroe,
Phys. Rev. Lett. \textbf{93}, 090410 (2004).


\bibitem{5}  L.-M. Duan and R. Raussendorf, Phys. Rev. Lett. 95, 080503
(2005).

\bibitem{6}  S. D. Barrett, P. Kok, Phys. Rev. A\ 71, 060310(R) (2005).


\bibitem{7}  L.-M. Duan, B. Wang, J. Kimble, Phys. Rev. A 72, 022320 (2005).


\bibitem{8}  Y. L. Lim et al., Phys. Rev. A 73, 012304 (2006); Y. L. Lim, A.
Beige, L. C. Kwek, Phys. Rev. Lett. 95, 030505 (2005).


\bibitem{exp}  M. J. Madsen et al., quant-ph/0603258 .

\bibitem{exp2}  D. Stick et al., Nature Physics \textbf{2}, 36 (2006); D.
Kielpinski, C. Monroe, and D. J. Wineland, Nature 417, 709 (2002).

\bibitem{surfacetrap} S. Seidelin, et al., quant-ph/0601173 (2006).

\bibitem{deslauriers} L. Deslauriers, et al., quant-ph/0602003 (2006).

\bibitem{note1}  If an ultrafast laser pulse is not available, one can
replace it by two phase-locked narrow-band pulses which drives the D1
transitions $\left| F,m=0\right\rangle \rightarrow \left| F^{\prime
}+1,m=0\right\rangle $ and $\left| F+1,m=0\right\rangle \rightarrow \left|
F^{\prime },m=0\right\rangle $ ($F^{\prime }=F$), respectively.


\bibitem{note2}  For $^{111}$Cd$^{+}$ (or for any atoms with the nuclear
spin $I=1/2$), one can also drive the D2 line $S_{1/2}\rightarrow P_{3/2}$,
where the two corresponding hyperfine transitions are given by $\left|
F,m=0\right\rangle \rightarrow \left| F^{\prime },m=0\right\rangle $ and $%
\left| F+1,m=0\right\rangle \rightarrow \left| F^{\prime
}+1,m=0\right\rangle $ with $F^{\prime }=F+1$, see Ref. \cite{exp}.


\bibitem{note3}  For instance, for $^{133}$Cs atoms or $^{111}$Cd$^{+}$
ions, the hyperfine splitting is about $9$ GHz ($14$ GHz), while the natural
linewidth of the excited level (the inverse of the lifetime) is around $5$
MHz ($60$ MHz). In both cases, the condition is well satisfied.


\bibitem{10}  L.-M. Duan, B. B. Blinov, D. L. Moehring, C. Monroe, Quant.
Inf. Comp. \textbf{4}, 165-173 (2004).

\bibitem{CQED} J. McKeever, J. R. Buck, A. D. Boozer, and H. J.
Kimble, Phys. Rev. Lett. \textbf{92}, 143601 (2004); P. Maunz, et
al., \textit{Nature} \textbf{428}, 50 (2004); J. A. Sauer, et al.,
Phys. Rev. A\textbf{69}, 051804 (2004); G. Guthorlein, et
al.,Nature \textbf{414}, 49 (2001); A. Mundt, et al., Phys. Rev.
Lett. \textbf{89}, 103001 (2002).

\bibitem{note4} Note, however, that the noise model assumed in
Ref. \cite{6} is more restrictive. It assumes there that if a gate
fails, states of the two target qubits are not destroyed, but
instead, they are subject only to Z type of errors (without bit
flips from $\left|0\right\rangle$ to $\left|1\right\rangle$, for
instance). In our physical implementation, however, due to the
existence of $\pi$-decaly channels (which are alwyas there for any
realistic atoms), there is a significant probability of bit-flip
errors when a gate fails. So, in the case of a gate failure, one
needs to trace out the destroyed qubits. The noise model here is
exactly the same as what is assumed in Ref. \cite{5}. We can
follow the contruction there to prove scalalbility under this kind
of noise model.

\bibitem{11}  H.J. Briegel and R. Raussendorf, Phys. Rev. Lett. \textbf{86},
910 (2001); R. Raussendorf and H. J. Briegel, Phys. Rev. Lett. \textbf{86},
5188 (2001).
\end{thebibliography}
\end{document}